\documentclass[%
 aip,
 amsmath,amssymb,
 reprint,%
]{revtex4-1}

\usepackage{graphicx}
\usepackage{dcolumn}
\usepackage{bm}

\usepackage[utf8]{inputenc}
\usepackage{mathptmx}
\usepackage{etoolbox}
\usepackage[T1]{fontenc}
\usepackage{float}
\usepackage{esint}
\usepackage[tight]{subfigure}
\usepackage{verbatim}
\usepackage{xspace}
\usepackage{mathtools}
\usepackage{etoolbox}
\usepackage{epstopdf}
\usepackage{epsfig}
\usepackage{amsmath}
\usepackage{amssymb}
\usepackage{amsfonts}
\usepackage{mathptmx}
\usepackage{eucal}
\usepackage{color}
\usepackage{eucal}
\usepackage{color}
\usepackage[colorlinks=true,linktoc=all,linkcolor=blue,breaklinks=true, citecolor=blue,urlcolor=blue]{hyperref}
\usepackage{natbib}
\usepackage{siunitx}

\makeatletter
\def\@email#1#2{%
 \endgroup
 \patchcmd{\titleblock@produce}
  {\frontmatter@RRAPformat}
  {\frontmatter@RRAPformat{\produce@RRAP{*#1\href{mailto:#2}{#2}}}\frontmatter@RRAPformat}
  {}{}
}%
\makeatother
\begin{document}

\preprint{AIP/123-QED}

\title[Single charge transport in a fully superconducting SQUISET locally tuned by self-inductance effects]{Single charge transport in a fully superconducting SQUISET locally tuned by self-inductance effects}
\author{E. Enrico}
 \email{e.enrico@inrim.it}
\author{L. Croin}%
\affiliation{ 
INRIM, Istituto Nazionale di Ricerca Metrologica, Strada delle Cacce
91, I-10135 Torino, Italy
}%

\author{E. Strambini}
\author{F. Giazotto}
\affiliation{%
NEST, Istituto Nanoscienze-CNR and Scuola Normale Superiore, Piazza
S. Silvestro 12, Pisa I-56127, Italy
}%

\date{\today}

\begin{abstract}
We present a single-electron device for the manipulation of charge states via quantum interference in nanostructured electrodes. Via self-inductance effects, we induce two independent magnetic fluxes in the electrodes and we demonstrate sensitivity to single charge states and magnetic field at variable temperature. Moreover, our approach allows us to demonstrate local and independent control of the single-particle conductance between nano-engineered tunnel junctions in a fully-Superconducting Quantum Interference Single-Electron Transistor (SQUISET), thereby increasing the flexibility of our single-electron transistors. Our devices show a robust modulation of the current-to-flux transfer function via control currents, while exploiting the single-electron filling of a mesoscopic superconducting island. Further applications of the device concept to single-charge manipulation and magnetic-flux sensing are also discussed.
\end{abstract}

\maketitle

\begin{figure*}[htp]
\centering
  \includegraphics[width=1.8\columnwidth]{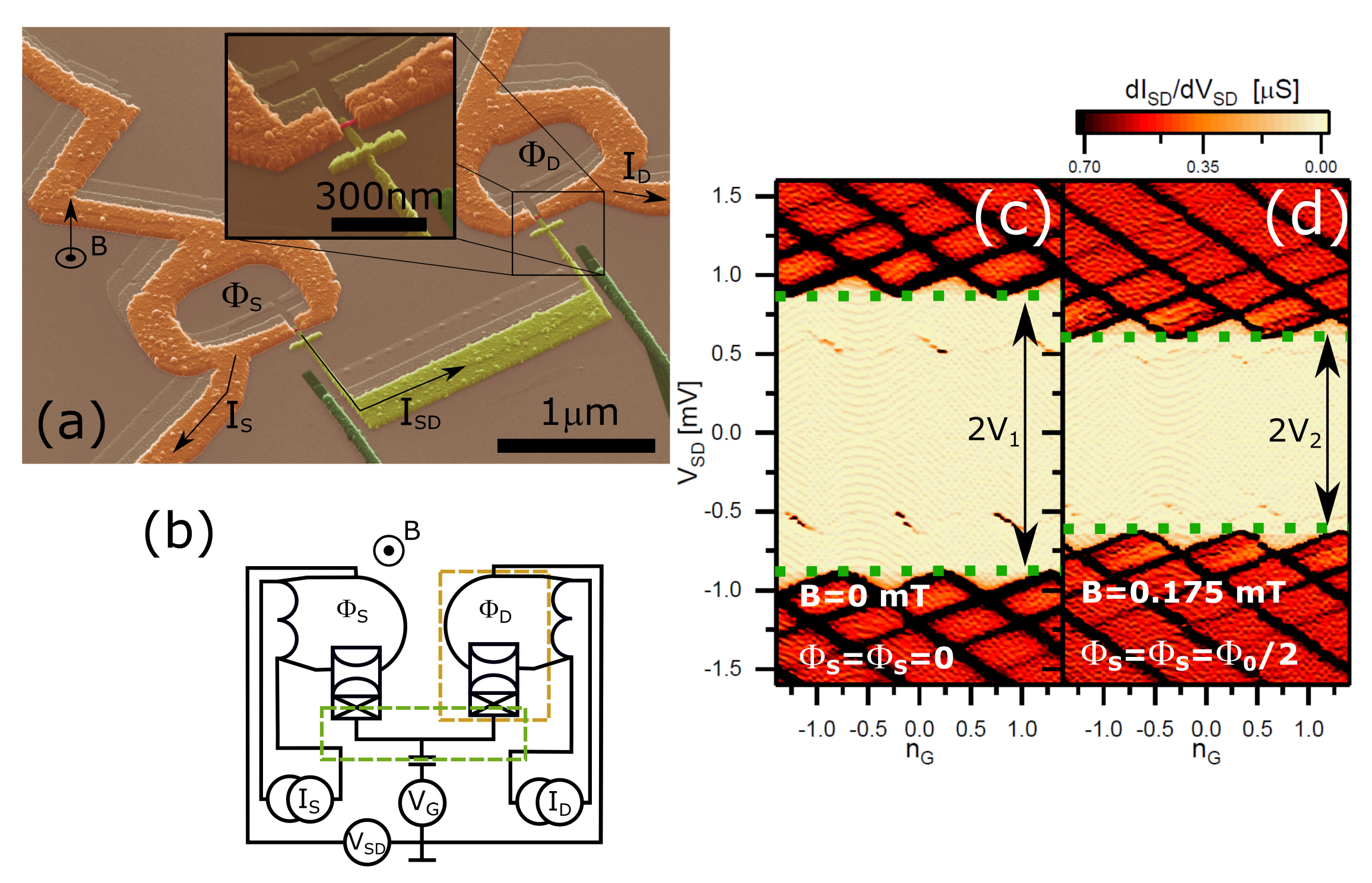}
  \caption{(a) Scanning electron micrography of a typical SQUISET device. The two currents paths ($I_{S}$
 and $I_{D}$) generating the two magnetic fluxes ($\Phi_S$ and $\Phi_D$) are indicated. The latter pierce the two superconducting loops of the source and drain electrodes (orange). A mesoscopic island (light green) is in tunnel contact with two superconducting nanowires (red) and it is capacitively coupled to a gate electrode (green). Few elements of the device can be attributed to fabrication or measurement details. In particular, the cross-like structure reported by the inset near the nanowire is its unavoidable duplicate coming from the shadow deposition technique used to fabricate the structure. Always in the shadow technique context, the fork-like structure of the gate electrodes guarantees a fixed distance between the island and the gate at every deposition angle. The structure with sharp angles composing the current biasing wires acts as mirrors for high frequencies components of the control parameters. (b) Circuital representation of the device. Two currents ($I_{S}$ and $I_{D}$) flow in two sections of the superconducting loops, while the device is entirely pierced by a uniform magnetic field ($B$). (c-d) Stability diagrams measured at $T=$\SI{30}{\milli\tesla} showing the differential conductance $dI_{SD}/dV_{SD}$ at different $n_G$ and $V_{SD}$ values when $\Phi_S=\Phi_D=0$ (c) and $\Phi_S=\Phi_D=\Phi_0/2$ (d). Here the magnetic fluxes are induced by the external magnetic field $B$. Black arrows indicating $2 V_1$ and $2 V_2$ represent the voltage region where the current is blocked by either the superconducting gaps of the island and the electrodes or the charging energy.}
  \label{fgr:Figure1}
\end{figure*}

Superconducting nanoelectronics has continuously grown in the last decades as a flexible and promising platform for the implementation of quantum-based sensors\cite{Giazotto2010,Ronzani2017,Dambrosio2015} and quantum-states manipulating circuits\cite{Uri2017,Strambini2016}, with particular attention to interference-based superconducting devices\cite{Virtanen2016} and mesoscopic structures where single charges play dominant roles\cite{Enrico2016,Enrico2017}. Different geometries can be easily combined with standard nanolithography techniques\cite{Fulton1987}, opening the field to complex and robust devices embedding multiple control lines and tunable working points in the parameters space.
As a consequence, superconducting nanoelectronics technology represents an exceptional research platform for condensed-matter quantum physics experiments as well as for scalable quantum computing\cite{Clarke2008} and photonics applications\cite{GU2017}. 

Normal-metal\cite{Pothier1992}, hybrid\cite{Pekola2007} or fully-superconducting\cite{Averin1992} single-electron devices - fabricated by shadow-mask technique \cite{Fulton1987} - have been so far one of the research topics where nanofabrication technology excelled, leading to device concepts where the detection of charge states approaching their coherent superposition\cite{Nakamura1999} has been routinely reached.
While rather complex single-electron systems based on local electrical gating have been demonstrated\cite{Martinis1994}, the on-chip tunability of their electrodes carriers population has been limited to the semiconductor nanowires\cite{Hollosy2015} and the 2D-electron-gas based technologies\cite{Giblin2012}, where clear manipulation of Coulomb blockade effects has only been allowed via strong electric fields.

Nano-engineered superconducting electrodes\cite{Enrico2017} introduce an alternative control parameter, the magnetic flux, that can act on the population of quasiparticles charge carriers\cite{Manninen1999} via quantum interference\cite{Giazotto2010}. Short metallic nanowires have been embedded in superconducting loops \cite{Enrico2016} leaving enough space to be coupled to a Coulombic island through mesoscopic tunnel junctions. 
The present technology, which is mostly based on aluminum tunnel junctions, is then further extended by an unprecedented level of control and flexibility offered by localized magnetic fluxes.   
Various approaches exploiting these phenomena demonstrated state-of-the-art magnetic flux sensing capabilities\cite{Ligato2017,Ronzani2017,Dambrosio2015} and single charges states manipulation\cite{Enrico2017} but still lack for on-chip control.

Here we demonstrate that two local magnetic fluxes can be used to manipulate the electrodes density of states of a fully superconducting SQUISET and to efficiently modify its electron transport properties. In particular, we show how the typical Coulomb energy of the island can be controlled by the quasiparticle spectra of the source and drain electrodes by exploiting self-inductance effects.

\begin{figure}[htp]
  \includegraphics[width=1\columnwidth]{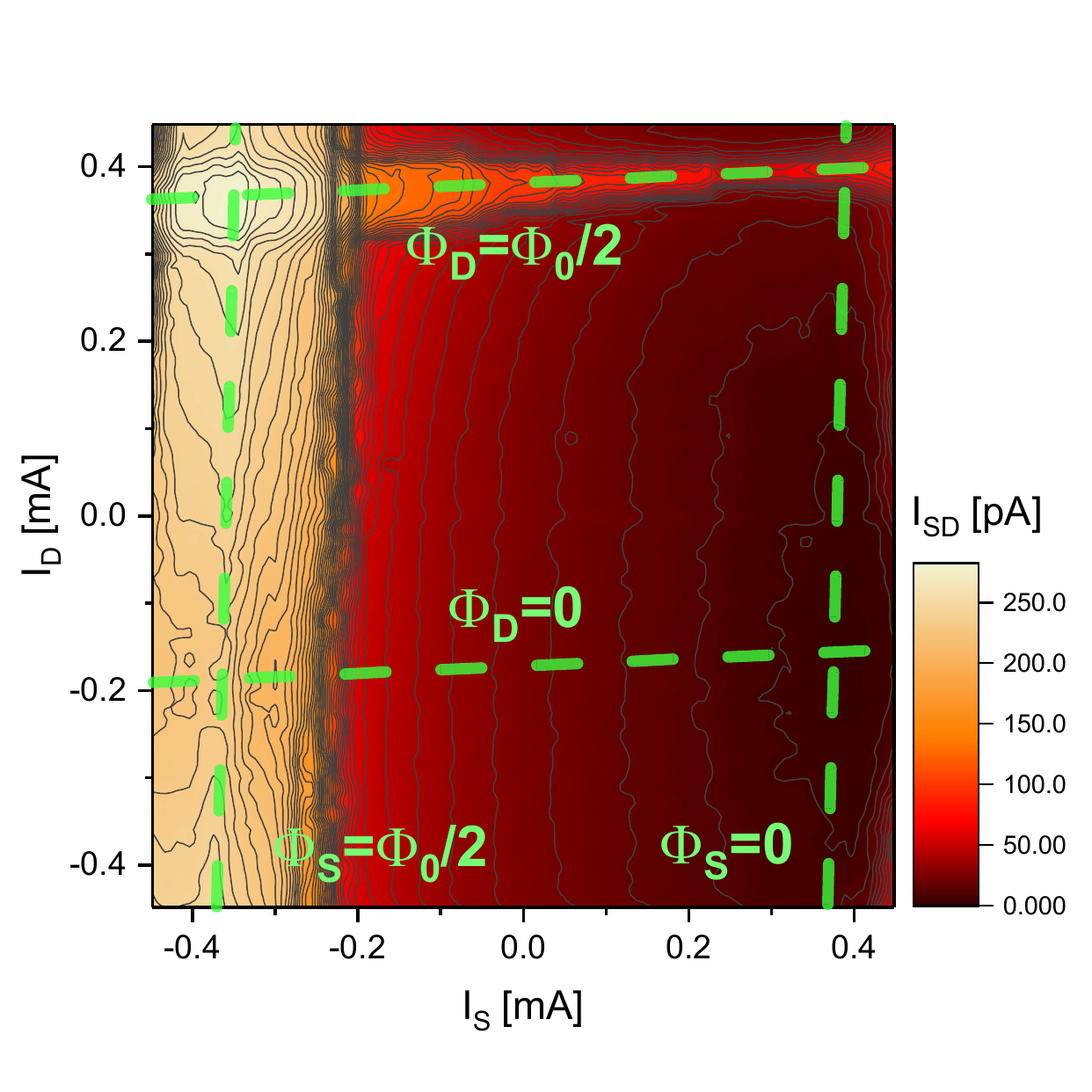}
  \caption{Contour plot of the source-drain current at fixed bias voltage ($V_{SD}=V_1$) versus on-chip control currents ($I_{S}$ and $I_{D}$). $B=$\SI{0.0875}{\milli\tesla} is applied leading to the condition $\Phi_B=\Phi_0/4$, and the device temperature is set to be $T=$\SI{700}{\milli\kelvin}.}
  \label{fgr:Figure2}
\end{figure}

\begin{figure*}[htp]
\centering
  \includegraphics[width=1.8\columnwidth]{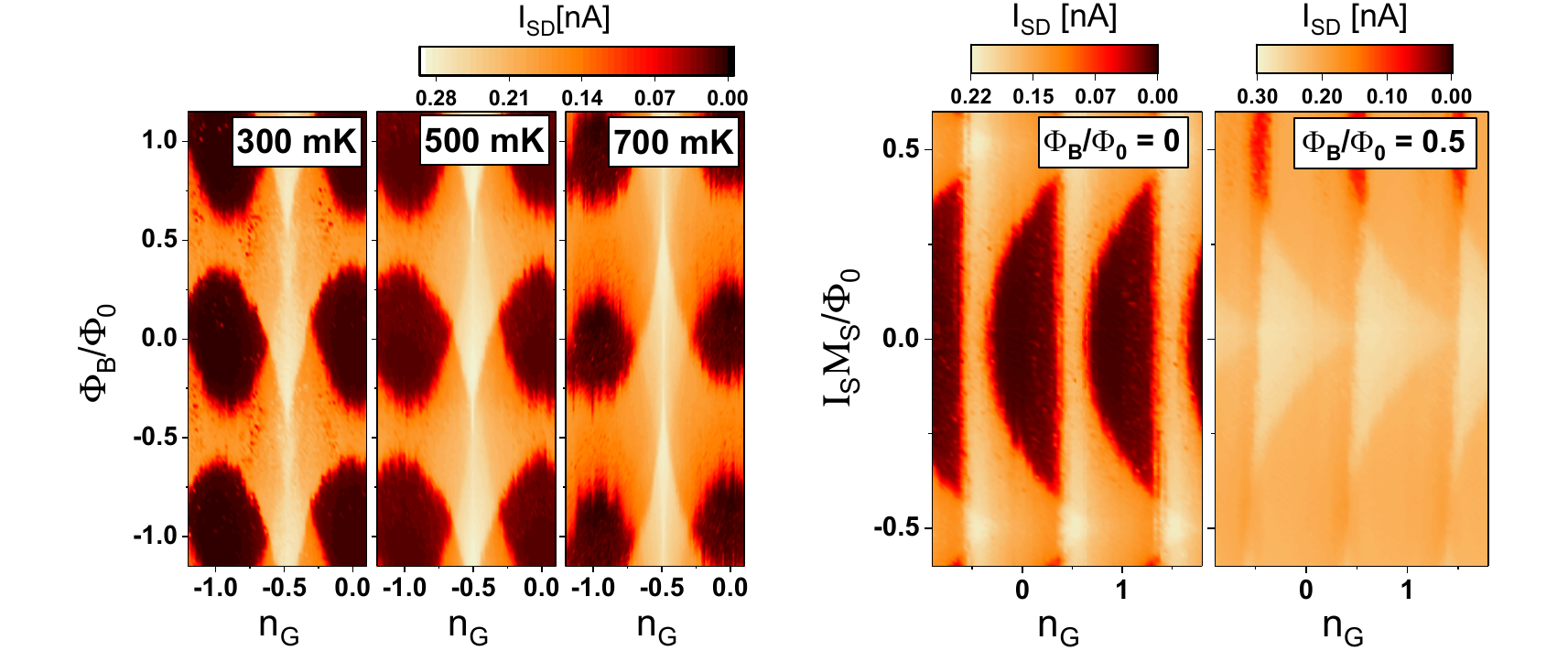}
  \caption{(a) Source-drain current ($I_{SD}$) versus normalized gate voltage ($n_G$) and magnetic flux ($\Phi_B$) at different bath temperatures. (b) Source-drain current in magnetic flux ($I_{S}M_{S}$) and electrical $n_G$ local gating condition at fixed temperature ($T=$\SI{700}{\milli\kelvin}) having superimposed a $\Phi_B=0$ and $\Phi_B=\Phi_0/2$ magnetic flux offset. All the measurement have been performed at the fixed bias condition $V_{SD}=V_1$.}
  \label{fgr:Figure3}
\end{figure*}

A prototypical device is depicted in Figure \ref{fgr:Figure1}. 
A superconducting island is connected to the source and drain electrodes via tunnel junctions. Both source and drain consist of a superconducting nanowire embedded in a superconducting loop. Each ring has two contact pads for the injection of the source-drain current and the currents for the independent control of the fluxes.  
The entire structure is realized via three-angle-deposition ($42^\circ$/$20^\circ$/$0^\circ$) of aluminum (\SI{15}{\nano\metre}/\SI{20}{\nano\metre}/\SI{100}{\nano\metre}) through a suspended mask on a Si/SiO2 (\SI{300}{\nano\metre} thick oxide) substrate (see Figure \ref{fgr:Figure1}a). The  polymeric mask  has been obtained via electron beam lithography, whereas thin films deposition has been performed via electron beam evaporation. Tunnel junctions were created between the first and the second deposition step by oxygen exposure (\SI{5e-2}{\milli\bar} for \SI{5}{\minute}). One of the tunnel junction across the nanowire and the island is visible in the inset of Figure \ref{fgr:Figure1}a. The device configuration defines three main current path $I_{S}$, $I_{D}$ and $I_{SD}$. The first two act as control currents flowing along parts of the source and drain loops, while the last is the effective current flowing through the Coulombic island (Figure \ref{fgr:Figure1}b). The entire chip is pierced by an uniform magnetic field, $B$, generated by an external magnet inducing a flux $\Phi_B=A*B$ in both the identical loops of area $A$. The combined effect of $B$ and the local currents gives rise to two magnetic fluxes at the source ad drain loops, $\Phi_S = \Phi_B + M_S*I_S+m_S*I_D$ and $\Phi_D = \Phi_B + M_D*I_S+m_D*I_D$ , respectively. $M_S$ and $M_D$ are the self-inductances while $m_S$ and $m_D$ are the mutual inductances between opposite loop.
The electrodes are biased via an external voltage source ($V_{SD}$), and the island is exposed to a control electric field via a capacitively-coupled gate that induces $n_G=C_GV_G/e$ quantized charges, being $C_G$ the gate-island capacitance, $V_G$ the gate voltage and $e$ the electron charge.
This device architecture is designed to act essentially as a fully superconducting single electron transistor\cite{Nakamura1995,Nakamura1996} with two identical tunnel junctions (total series resistance $R_T\approx$\SI{1.75}{\mega\ohm}). 
In the absence of a magnetic field, this is confirmed by the differential conductance stability diagram in Figure \ref{fgr:Figure1}c clearly showing the effect of the charging energy, evaluated to be $E_C=$\SI{75}{\micro\electronvolt} from the Coulomb diamonds and confirmed by the Josephson-quasiparticle peaks (JQPs)\cite{Nakamura1996,Korotkov1996,Fulton1989,Brink1991,Averin1992}.
In particular, dark and sharp JQPs conductance peaks clearly visible in the blocked region of Figures \ref{fgr:Figure1}c and \ref{fgr:Figure1}d result to be unaffected by the small magnetic field applied since they depend on the island superconducting gap $\Delta_I$ and $E_C$ only. 
Therefore, from the JQPs we have estimated $\Delta_I\approx$\SI{216}{\micro\electronvolt}.
When the SQUISET is uniformly pierced by $B$, the condition $\Phi_S=\Phi_D=\Phi_B=\Phi_0/2$ can be reached, as show in Figure \ref{fgr:Figure1}d, and the superconducting gaps of the the two nanowires are reduced to their minimum via quantum interference. This effect can be appreciated by the reduction of the voltage threshold separating the conducting region, where the transport is dominated by quasiparticles tunneling and not JQP cycles, respect the blocked one ($V_1 = 2\Delta_{I}+\Delta_{S,0}+\Delta_{D,0}$ in Figure \ref{fgr:Figure1}c and $V_2= 2\Delta_{I}+\Delta_{S,1/2}+\Delta_{D,1/2}$ in Figure \ref{fgr:Figure1}d).
It's worth mentioning here that $V_1$ and $V_2$ have been selected as reference thresholds, for which the independence by the charging energy $E_C$ is guaranteed by their position respect the coulomb diamonds.
From there, the zero magnetic field and the $\Phi_0/2$ superconducting gaps of the electrodes have been deduced ($\Delta_{S,0}=\Delta_{D,0}\approx$\SI{235}{\micro\electronvolt} and $\Delta_{S,1/2}=\Delta_{D,1/2}\approx$\SI{84}{\micro\electronvolt}).
\begin{figure*}
\centering
  \includegraphics[width=1.8\columnwidth]{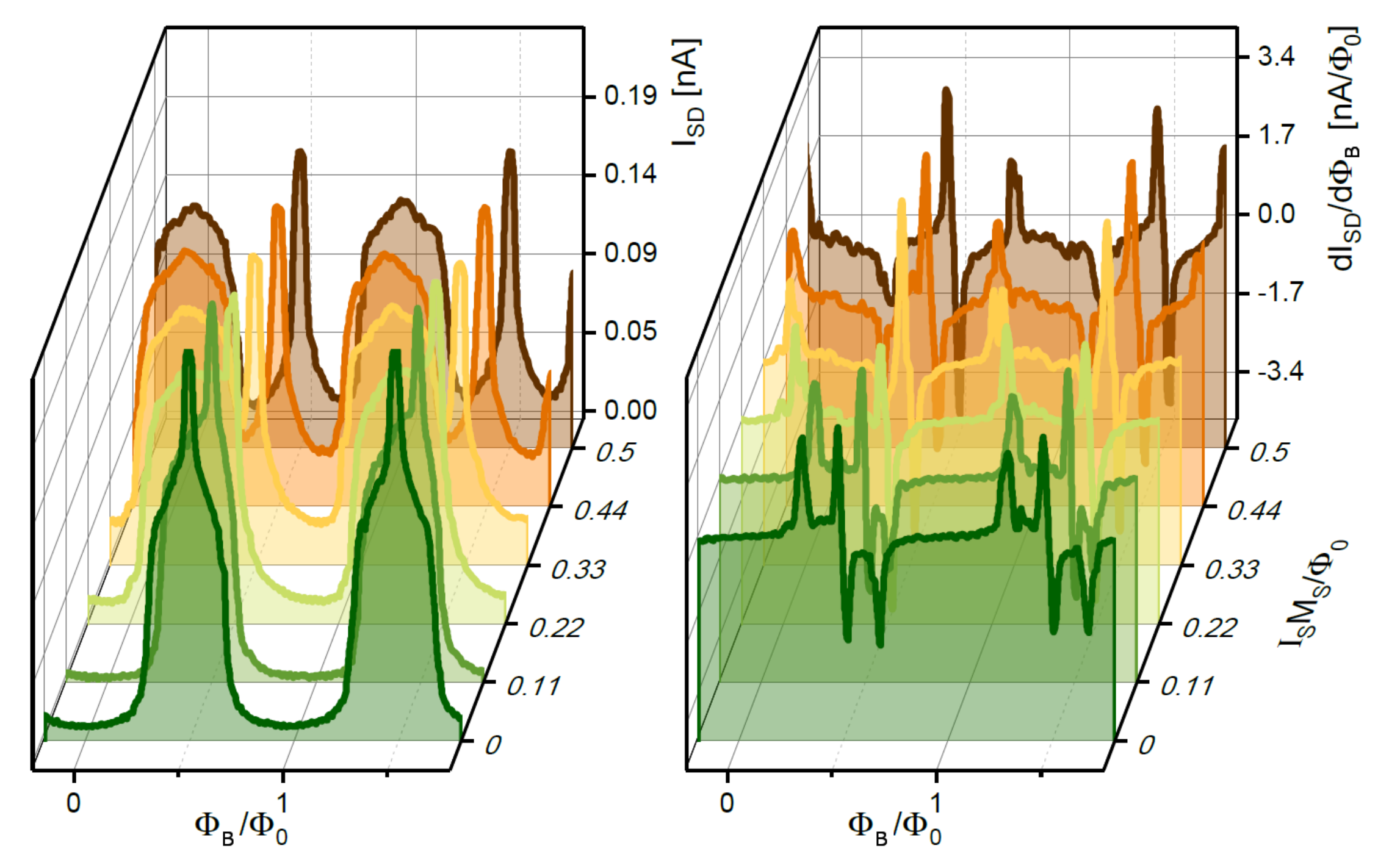}
  \caption{(a) Flux-modulated current ($I_{SD}$) at different values of $I_{S}$. (b) Flux-to-current ($dI_{SD}/d\Phi_B$) transfer function obtained from numerical derivation of the curves in (a). All the measurement were performed at $n_G=0$, $T=$\SI{700}{\milli\kelvin} and $V_{SD}=V_1$.}
  \label{fgr:Figure4}
\end{figure*}
The effect of local magnetic flux biasing via $I_{S}$ and $I_{D}$ is shown in Figure \ref{fgr:Figure2}, where the source-drain current $I_{SD}$ is monitored at fixed bias $V_{SD}=V_1$ as a functions of the currents flowing in the loops.
The non-symmetrical behavior shown in Figure \ref{fgr:Figure2} suggests an asymmetry in the dynamical conductance of the two tunnel junction involved. 
From the analysis of maxima and minima fitted positions in this diagram, represented by quasi-orthogonal light green dashed lines, it is possible to observe and quantify the effect of the self-inductances, giving $M_S=0.69 \Phi_0/$\si{\milli\ampere} and $M_D=0.87 \Phi_0/$\si{\milli\ampere}. From these estimates, the cross-influence of the flux control lines turns out to be almost negligible ($m_{S,D}<0.05 \Phi_0/$\si{\milli\ampere}) and the electrodes quasiparticles density of states results to be almost independently tunable by $I_{S}$ and $I_{D}$, respectively.
In order to further investigate the effect of an independent flux biasing via local effects, we have performed a temperature series measurements that confirm the single-charge sensitivity of our device up to $T=$\SI{700}{\milli\kelvin} (Figure \ref{fgr:Figure3}(a)). There, a symmetrical magnetic flux biasing condition via $B$ respects the periodical modulation of the source-drain current when the device is biased at $V_{SD}=V_1$. It's worth mentioning that experimental data reported in Figure \ref{fgr:Figure3} are affected by unavoidable background charge as commonly happens in single-charge sensitive devices. These offsets where removed by referring to the $n_G$ normalized quantized induced charge value, whose zero is centered in one of the $I_{SD}$ minima. By exploiting this evidence, we proceeded investigating the effect of local magnetic and electrical gating at $T=$\SI{700}{\milli\kelvin} and two different magnetic field leading to $\Phi_B=0$ and $\Phi_B=\Phi_0/2$ (see Figure \ref{fgr:Figure3}(b)). There, the periodical and asymmetrical dependence of $I_{SD}$ on $n_G$ reflects the unbalanced condition of the electrodes superconducting gaps that affects the conductances of the tunnel junctions, with clear similarities to the behavior reported in Figure \ref{fgr:Figure3}(a).  Triangular regions in the $n_G$-$I_{S}$ plane, corresponding to maximum $I_{SD}$ current, are shifted and expanded from the $I_{S}M_S=\Phi_{0}/2$ condition (when no external magnetic field is applied) to $I_{S}=0$ when a uniform magnetic flux offset is introduced ($\Phi_B=\Phi_0/2$). Analogously, semi-circular regions corresponding to blockaded regions of almost zero $I_{SD}$ current are shrunk and shifted around the $I_{S}M_S=\Phi_0/2$ condition.  
The mechanism of unbalanced response to magnetic field is analyzed in detail in Figure \ref{fgr:Figure4}(a), where we report the evolution of the flux-modulated current ($I_{SD}$) at different $I_{S}$. $I_{SD}$ presents sharp and periodic peaks on top of broader peaks. These latter are controlled by $I_S$ which induces their gradual separation. Yet, the sharp structures depends only on $I_D$, while their sharpness stems from the asymmetry existing between local conductances of the source and drain tunnel junctions \cite{Borchia}.
The S'ISIS'' structure of our device expresses here strong asymmetrical behavior respect the the symmetrical geometry, simply due to the local action of unbalancing flux given by self-inductance effects.  
Sharp peaks at $\Phi_B=\Phi_0/2$ are independent respect the current $I_{S}$ and can be attributed to the island-drain junction, confirming the negligible correlation between the two flux control lines of our device. The wider plateau of $I_{SD}$ can be shifted along the $\Phi_B$ axis at will by acting on the $I_{S}$ current. These plateau are clearly wider respect the sharp peaks of the island-drain junction due to the asymmetric voltage bias of the circuit (see Figure \ref{fgr:Figure1}(b)).
In order to quantify the flux-to-current transfer function we show in Figure \ref{fgr:Figure4}(b) the numerical derivative of $I_{SD}$ respect to  $\Phi_B$. Double peaked transfer functions reflect the role of the two different superconducting gaps, moreover the effect of the flux bias via $I_{S}$ can be exploited to further increment the responsiveness of our device to magnetic flux variation. As an example, when $I_{S}M_S=0.22\Phi_0$ the two negative peaks collapse in one and effectively enhance the transfer function from $|dI_{SD}/d\Phi_B|\approx$\SI{1.6}{\nano\ampere}$/\Phi_0$ to $|dI_{SD}/d\Phi_B|\approx$\SI{3.2}{\nano\ampere}$/\Phi_0$. Eventually, non negative responsiveness can be induced around $\Phi_B=0$ when $0.22\Phi_0<I_{S}M_S<0.33\Phi_0$. 
The high responsiveness of the SQUISET to magnetic field is a consequence of the Coulombic island enhancing the transfer function by acting as an energy filter\cite{Bhadrachalam2014} for the intermediate charge states involved in the transport processes. This flexible configuration confirms potential application of dynamical conductance-enhanced sensitivity to magnetic field variations in double-junction system embedding quantum interference based electrodes.

In summary, we have reported the fabrication and characterization of a fully-superconducting SQUISET demonstrating local manipulation of charge and magnetic flux sensing via independent current and voltage control lines. We discuss in detail the dependencies on external magnetic field, gate voltage, flux bias currents and temperature, which is possible due to the multiple-electrodes design of the device. On one side, this proof-of-concept device opens up to an unprecedented tools to superconducting charge control, with quantum interference based nanostructured electrodes, to be used in quantum electronics\cite{Uri2017} and metrology\cite{PekolaRevModPhys2013,Giblin2012,Kaneko2016}. Moreover, straightforward integration with present quantum technologies\cite{Clarke2008} based on aluminum nanostructures is worth considering.  On the other side, the enhanced and flexible sensitivity to magnetic fields envisage our device concept for the implementation of energy-filtered\cite{Bhadrachalam2014} single charge magnetometers.
\begin{acknowledgments}
The authors gratefully acknowledge Compagnia di San Paolo for financial support to NanoFacility Piemonte at INRIM. They also thank G. Amato, L. Callegaro and I. Mendes for helpful discussions. This work was supported by the INRiM "IBC-QuBit" - Seed Project.
E. S. and F. G. acknowledge financial support from the ERC grant agreement no. 615187 - COMANCHE. F.G. acknowledge the Royal Society though the International Exchanges between the UK and Italy (grant IESR3 170054). The work of F. G. was partially funded by the Tuscany
Region under the FARFAS 2014 project SCIADRO. 
\end{acknowledgments}
\section*{Data Availability Statement}
The data that support the findings of this study are available from the
corresponding author upon reasonable request.

\bibliography{aipsamp}

\end{document}